\documentclass[twocolumn,prl,superscriptaddress,showpacs]{revtex4-1} 

\usepackage{graphicx} 
\usepackage{natbib} 
\usepackage[usenames,dvipsnames]{color} 
\usepackage{amsmath} 
\usepackage{amssymb} 
\usepackage{soul} 
\usepackage{ifthen} 
\usepackage{array}

\newcommand{\dg}{^\dagger}

\newcommand{\rt}[1]{\sqrt{#1}}
\newcommand{\bra}[1]{\langle{#1}|}
\newcommand{\ket}[1]{|{#1}\rangle}

\def\l{\left}
\def\r{\right}
\def\be#1\ee{\begin{equation}#1\end{equation}}
\def\ba#1\ea{\begin{align}#1\end{align}}
\def\bg#1\eg{\begin{gather}#1\end{gather}}
\def\t{\text}

\newcommand{\abs}[1]{\lvert#1\rvert}

\def\shownote{0} 
\newcommand{\note}[1]{\ifthenelse{\shownote=1}{\textcolor{Red}{[[#1]]}}{}}
\newcommand{\nt}[1]{\ifthenelse{\shownote=1}{\textcolor{Red}{[[#1]]}}{}}
\def\shownoteforauthors{0} 
\newcommand{\nta}[1]{\ifthenelse{\shownoteforauthors=1}{\textcolor{Orange}{[[#1]]}}{}}
\def\showaddmat{0} 
\newcommand{\addmat}[1]{\ifthenelse{\showaddmat=1}{\textcolor{Gray}{[[#1]]}}{}}

\begin{document}

\title{Dynamics of parametric fluctuations induced by quasiparticle tunneling in superconducting flux qubits }

\author{M. Bal}
\affiliation{Institute for Quantum Computing, Department of Physics and Astronomy, and Waterloo Institute for Nanotechnology, University of Waterloo, Waterloo, ON, Canada N2L 3G1}
\affiliation{TUBITAK Marmara Research Centre, Materials Institute, P.O. Box 21, 41470 Gebze, Kocaeli, Turkey}

\author{M. H. Ansari}
\affiliation{Institute for Quantum Computing, Department of Physics and Astronomy, and Waterloo Institute for Nanotechnology, University of Waterloo, Waterloo, ON, Canada N2L 3G1}
\affiliation{Kavli Institute of Nanoscience, Delft University of Technology, P.O. Box 5046, 2600 GA Delft, The Netherlands}

\author{J.-L. Orgiazzi}
\affiliation{Institute for Quantum Computing, Department of Electrical and Computer Engineering, and Waterloo Institute for Nanotechnology,
University of Waterloo, Waterloo, ON, Canada N2L 3G1}

\author{R. M. Lutchyn}
\affiliation{Station Q, Microsoft Research, Santa Barbara, CA 93106-6105}

\author{A. Lupascu \footnotemark[1] \footnotetext[1]{Corresponding author: alupascu@uwaterloo.ca}}
\affiliation{Institute for Quantum Computing, Department of Physics
and Astronomy, and Waterloo Institute for Nanotechnology, University
of Waterloo, Waterloo, ON, Canada N2L 3G1}

\date{ \today}

\begin{abstract}
We present experiments on the dynamics of a two-state parametric fluctuator in a superconducting flux qubit. In spectroscopic measurements, the fluctuator manifests itself as a doublet line.
When the qubit is excited in resonance with one of the two doublet lines, the correlation of readout results exhibits an exponential time decay which provides a measure of the fluctuator transition rate. The rate increases with temperature in the interval 40 to 158 mK. Based on the magnitude of the transition rate and the doublet line splitting we conclude that the fluctuation is induced by quasiparticle tunneling. These results demonstrate the importance of considering quasiparticles as a source of decoherence in flux qubits.

\end{abstract}
\maketitle

Superconducting qubits are one of the most promising class of candidate systems for the implementation of a quantum information processor~\cite{Devoret_2013_SupCircuits,clarke_2008_rev-sup-qb}. Developments in this field depend critically on the qubit quantum coherence times. Significant advances on improving coherence times were made recently by the introduction of qubits in three dimensional cavities~\cite{paik_2011_3Dtransmon,rigetti_2012_0p1msqubit} as well as by optimization of the design of qubits in a planar geometry~\cite{chow_2012_ApproachFaultTolerant,sandberg_2013_LongLivedTransmon}. Despite these advances, many features of decoherence in superconducting systems are only partially understood.

Decoherence of superconducting qubits is induced by the noise generated in a complex solid-state environment. Further understanding of the sources of decoherence requires measuring the properties of the noise, which is done most effectively by using the qubits themselves. This approach requires the measurement of qubit evolution combined with the control of the susceptibility to different noise channels. Besides the benefits for quantum information, using superconducting qubits to measure noise brings new and exciting opportunities to experimentally investigate the physics of noise in mesoscopic systems. As an example, qubits were used to perform detailed measurements of the spectral density of flux noise over a wide frequency range~\cite{bylander_2011_noisePCQ,slichter_2012_FluxNoisecQED}, considerably expanding the spectral interval accessible by superconducting quantum interference devices (SQUID) measurements~\cite{wellstood_1987_dcsquidnoise}.

In this letter, we present experiments in which we probe the dynamics of a two-state fluctuator (TSF) coupled to a superconducting flux qubit. TSFs are a generic type of noise, observed in many mesoscopic systems, with examples including charge~\cite{Zimmerli_1992_ChargeNoise}, flux~\cite{gail_1998_VorticesPinningAndDepinning}, and critical current fluctuators~\cite{eroms_2006_1}. In most of these experiments, TSFs are characterized using classical detectors, such as single-electron transistors~\cite{Zimmerli_1992_ChargeNoise} or SQUIDs~\cite{gail_1998_VorticesPinningAndDepinning}. In this letter, we present a method to determine the time scales of a TSF which relies on conditional excitation and measurement of a qubit. Based on the parametric change of the qubit frequency and the measurement of the TSF time scales, we conclude that the TSF origin is tunneling of quasiparticles through the Josephson junctions forming the qubit. Our results provide new insight into the decoherence of flux-type superconducting qubits.

\begin{figure}[!]
\includegraphics[width=3.4in]{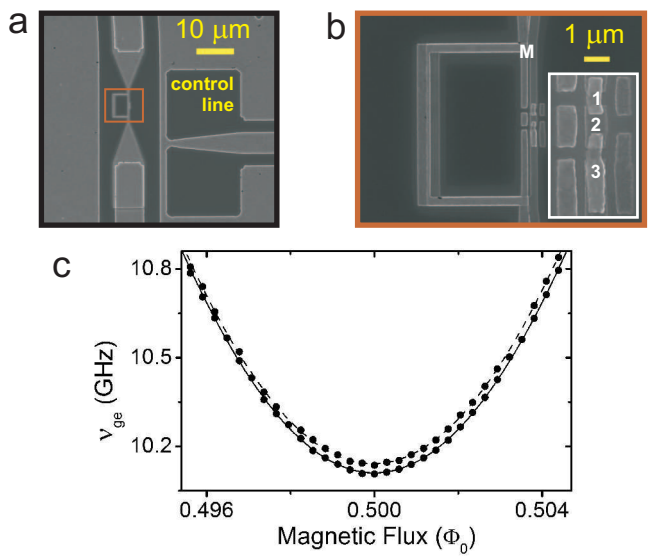}
\caption{\label{fig1} \textbf{(a)} Scanning electron microscope image of a device nominally identical to that used in this work. The qubit, which is the loop inside the rectangle, is embedded into a CPW resonator. The CPW on the right is used for qubit control. \textbf{(b)} Zoom of the region indicated by the rectangle in (a). The large superconducting electrode is labeled $M$, and the qubit islands, shown in the inset, are labeled by 1 to 3. \textbf{(c)}  Qubit spectroscopy measurements. The continuous (dashed) line is a fit of the low (high) energy doublet line frequency with the standard PCQ model; the fit yields $I_{p}=138.6\,\t{nA}$ and $\Delta=10.11\,\t{GHz}$ ($I_{p}=139.0\,\t{nA}$ and $\Delta=10.14\,\t{GHz}$).}
\end{figure}

The qubit used in our experiments is a persistent current qubit (PCQ)~\cite{mooij_1999_1}, a flux-type superconducting qubit. This qubit is coupled to a superconducting coplanar waveguide (CPW) resonator, in a circuit-quantum electrodynamics type architecture~\cite{Bal_2012_magnetometer,abdumalikov_2009_flux-qubit,niemczyk_2010_cqedstrong,jerger_2011_MultiplexPCQ}. The CPW resonator, coupled inductively to the qubit (see Fig.~\ref{fig1}a), has a resonance frequency $\nu_{\t{res}} = 6.602$ GHz, significantly lower than the qubit transition frequency. The state of the qubit is measured by applying a microwave readout pulse of duration $T_\t{r}$ and frequency $\nu_{\t{r}} = \nu_{\t{res}}$ to the CPW resonator. After transmission through the CPW resonator, this pulse is down-converted and its average, the homodyne voltage $V_\t{H}$, is kept as a qubit measurement record. The qubit is controlled using microwave signals sent through a separate CPW control line (Fig.~1a). The device is microfabricated on a silicon wafer, in a two-step process. In the first step, optical lithography and lift-off is used to define a $200$ nm thick aluminum layer containing the CPW resonator and the CPW control line. In the second step, electron-beam lithography followed by standard shadow evaporation of aluminum and lift-off is used to realize the PCQ. The PCQ consists of a superconducting ring interrupted
by four Josephson junctions, formed by two aluminum layers, with thicknesses $40$ nm and $65$ nm respectively, separated by a thin
in-situ grown aluminum oxide layer (see Fig.~\ref{fig1}b). All measurements are performed using a custom-designed probe in a dilution
refrigerator~\cite{Ong_2012_Probe}.

\begin{figure}[!]
\includegraphics[width=3.4in]{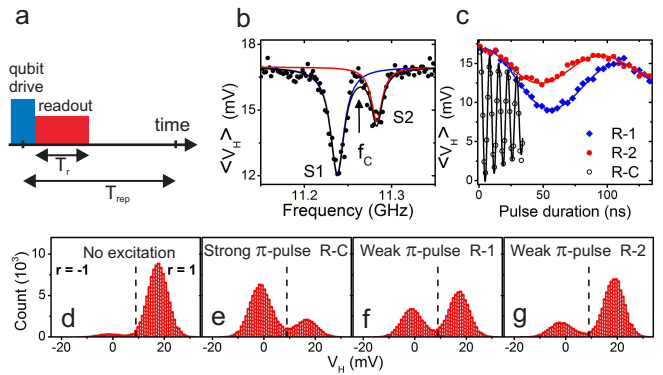}
\caption{\label{fig2} \textbf{(a)} Sequence used for qubit state preparation and readout. The readout time is $T_\t{r}=520\,\t{ns}$ and the repetition time is $T_{\t{rep}}=10\,\mu\t{s}$. \textbf{(b)} Spectroscopy of the qubit. The doublet lines are labeled S1 and S2. \textbf{(c)}Rabi oscillations for strong excitation at frequency $f_c$ in (b) (setting R-C) and weak excitation at lines S1 and S2 (settings R-1 and R-2 respectively).
\textbf{(d-g)} Histograms of the homodyne voltage values $V_\t{H}$ for no qubit excitation (d), and $\pi$ pulse excitation at settings R-C, R-1, and R-2 (e, f, and g respectively), for $10^5$ repetitions. The vertical dashed line at
$8.95$ mV indicates the position of the threshold used to separate the readout values labeled $r=1$ and $r=-1$.}
\end{figure}

We first characterize the qubit by performing spectroscopic measurements, with the results shown in Fig.~\ref{fig1}c. We expect to observe a dip in the homodyne voltage $V_H$ when the excitation frequency matches the ground to first excited state separation. However, our
spectroscopic measurements reveal a double, rather than a single, resonance line (see Fig.~\ref{fig2}b for a typical spectroscopy curve).

We rule out microscopic quantum two-level systems (TLSs) as the source of the observed doublet, as coupled TLSs produce an avoided crossing in the spectrum.~\cite{martinis_2005_1,lupascu_2010_1} Contrary to the case of TLSs coupled to a qubit, our observations portray a picture where there are two sets of parameters describing the qubit, resulting in two transition frequencies. We start by presenting our experimental results without any assumptions on the mechanism for the observed parametric change. Next, a method is developed to extract the time
scales associated with the parametric changes of the qubit frequency. In the last part of the paper we discuss the possible physical origin of these effects.

The observed spectroscopy doublet suggests a two-state fluctuator acting on the qubit. Indeed, let us assume that the two states of the TSF, labeled in the following as S1 and S2, result in two different qubit transition frequencies. In spectroscopy experiments, a point at one given frequency is obtained by averaging typically $10^{4}$ repetitions of a sequence, shown in Fig.~\ref{fig2}a, consisting of qubit excitation by a weak pulse followed by readout. If the average dwell time for each of the two TSF states is much longer than the sequence repetition time, yet much shorter than the time required to complete all the repetitions, we expect the average signal to display a resonance at both transition frequencies.

The hypothesis of a qubit transition frequency which changes between two values is further confirmed by the following experiments. We measured Rabi oscillations in three different settings: strong driving with a microwave frequency corresponding to the center of the doublet, with a Rabi frequency significantly larger than the doublet splitting (setting denoted R-C), and weak driving with a frequency corresponding to either doublet line, with a Rabi frequency much smaller than the doublet separation (settings denoted R-1 and R-2 respectively). The Rabi oscillation amplitude for setting R-C is approximately equal to the sum of the Rabi oscillation amplitudes for settings R-1 and R-2 (see Fig.~\ref{fig2}c). This is consistent with full excitation of the qubit for R-C, as opposed to partial, TSF-state dependent, excitation for R-1 and R-2. This conclusion is further supported by a measurement of readout homodyne voltage histograms, for no qubit excitation (Fig.~\ref{fig2}d) and $\pi$ pulses for settings R-C, R-1, and R-2 respectively (Fig.~\ref{fig2}e-g). The histograms show a bimodal distribution; a threshold is used to separate intervals corresponding to the ground(g) and excited(e) states of the qubit, labeled as $r=1$ and $r=-1$ respectively. The weight of the $r=-1$ part for R-C has a value close to the sum of the $r=-1$ weights for R-1 and R-2.

\begin{figure}[!]
\includegraphics[width=3.0in]{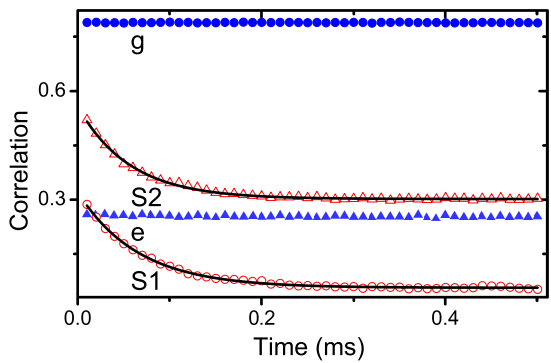}
\caption{\label{fig3}  Correlation of readout results versus time for qubit ground state (g, full circles), excited state (e, full triangles), and partial excitation at lines S1 (empty circles) and S2 (empty triangles). The solid lines are fits with exponential decay functions.}
\end{figure}

To unveil the dynamics of the TSF, we perform an experiment in which we repeat a sequence formed of qubit excitation with an R-1 $\pi$ pulse followed by measurement. Let us first consider the ideal case of perfect Rabi rotations and readout fidelity. If the TSF is in state S1/S2 during the excitation pulse, then, after the Rabi pulse, the qubit is in state e/g, and therefore the readout result is $r=-1$/$r=1$ respectively. The qubit readout result is in a one-to-one correspondence with the TSF state, and therefore it allows probing of the TSF dynamics. However, due to decoherence and nonideal pulses and readout, this correspondence is not exact, yet statistical correlations exist between the TSF state and the readout result. Therefore, we analyze the experiment based on the \emph{correlation} of measurement results. We introduce:
\be \ c_j\equiv \frac{1}{N-j} \sum_{i=1} ^{N-j} r_{i}r_{i+j} \ee
where $r_i$, with $1\leq i\leq N$, is the $i$th result in a series of $N$ repetitions. The correlation is shown in Fig.~\ref{fig3} as a function of the time $jT_{\t{rep}}$, with $T_{\t{rep}}$ the repetition time. The correlation decays exponentially with a rate $\Gamma_{c1}$, a signature of transitions between the states of the TSF over the corresponding time scale. We can quantitatively relate the observed decay function to the TSF dynamics, if we assume that the dynamics of the TSF is described by a random telegraph noise process. With transition rates between the TSF states denoted by $\gamma_{S1\rightarrow S2}$ and $\gamma_{S2\rightarrow S1}$, the correlator is expected to decay exponentially with a rate $\Gamma_{c1}=\gamma_{S1\rightarrow S2} + \gamma_{S2\rightarrow S1}$. When the qubit is excited with an R-2, instead of R-1, $\pi$ pulse, an exponential decay is observed as well, with a rate $\Gamma_{c2}$ close to $\Gamma_{c1}$. This result is consistent with the assumption of telegraph noise: $\Gamma_{c2}=\gamma_{S2\rightarrow S1} + \gamma_{S1\rightarrow S2}=\Gamma_{c1}$. We also find that for no excitation of the qubit or excitation using a R-C $\pi$ pulse the correlation function has no time dependence, consistent with the qubit state being independent of the TSF state for these cases.

We now discuss the possible physical origin of a TSF consistent with our observations. We consider first a TSF which acts on the qubit via magnetic flux. For a flux qubit, the transition frequency $\nu_{\t{ge}}$ depends on the magnetic flux $\Phi$ as $\nu_{\t{ge}}(\Phi)=\rt{\Delta^2+\left( \frac{2I_\t{p}}{h}\left(\Phi-\frac{\Phi_0}{2} \right) \right)^2}$~\cite{orlando_1999_1}, where $\Phi_0$ is the flux quantum and $\Delta$ and $I_p$ are parameters which depend on the qubit junctions. It is not possible to explain the spectroscopic peak positions shown in Fig.~\ref{fig1}c based on two sets of transitions frequencies, given by $\nu_{\t{ge}}(\Phi+\Phi_{\t{S1}})$ and $\nu_{\t{ge}}(\Phi+\Phi_{\t{S2}})$, with $\Phi$ the applied magnetic flux and $\Phi_{\t{S1}}$ and $\Phi_{\t{S2}}$ the magnetic flux induced by the TSF in the states S1 and S2 respectively. We consider next the possibility of a TSF coupled to the qubit via electric field. For each island $i$ in the circuit ($i=\overline{1,3}$, see Fig.~\ref{fig1}), we model the effect of electric fields by a voltage source $V_{gi}$ coupled to the island via a capacitance $C_{gi}$, generalizing the model in~\cite{orlando_1999_1}. The Hamiltonian acquires a dependence on the gate charges $n_{gi}=C_{gi} V_{gi}/2e$.  In mesoscopic devices, the gate charge displays random fluctuations of microscopic origin. In some cases, the charge noise is found to contain a significant random telegraph noise component~\cite{Zimmerli_1992_ChargeNoise}. The transition frequency of the qubit, $\nu_{ge}(\Phi,n_{g1},n_{g2},n_{g3})$, may have a significant dependence on the gate charges $n_{\t{g}i}$ associated with the three qubit islands. For given values of the gate charges, the qubit spectrum can be well approximated by the relation $\nu_{\t{ge}}(\Phi)$ introduced above, with the parameters $I_{\t{p}}$ and $\Delta$ dependent on the gate charges. This is indicated by the fits in Fig.~\ref{fig1}c. The two values of $\Delta$ determined from the fit are different by 30~MHz. This difference is well within the range $\delta\Delta_{c}=207\,\t{MHz}$ of modulation of $\Delta$ by the variation of gate charges, calculated numerically.

\newcolumntype{x}[1]{%
>{\centering\hspace{0pt}}p{#1}}%

\begin{table}[t]
\caption{Summary of calculated charge modulation ($\delta \Delta _c$) and maximum observed doublet splitting for five qubit devices. For W37\_C2d\_Qb1 and W37\_C2d\_Qb3, no doublet is observed; we indicate the measured linewidth as an upper bound. The transition rate measurements presented in this paper are performed on sample W33\_B1b. \label{tab1}}
\begin{ruledtabular}
\begin{tabular}{p{2.4cm}|p{1.1cm}|p{1.8cm}|p{2.4cm}}
Sample & $E_J/E_c$ & Calculated $\delta\Delta_{c}$ (MHz) & Maximum observed splitting (MHz) \\ \hline
W33\_B3d & 1.5 & 1228 & 244 \\
W33\_B1d & 2.6 & 378 & 275 \\
W33\_B1b & 2.8 & 207 & 52 \\
W37\_C2d\_Qb1 & 10.7 & 0.083 & $<$ 0.79 \\
W37\_C2d\_Qb3 & 11.3 & 0.052 & $<$0.68
\end{tabular}
\end{ruledtabular}
\end{table}

We performed spectroscopy experiments on five qubits, fabricated using the same procedure and measured using very similar setups. The results are summarized in Table~\ref{tab1}. We indicate the Josephson energy $E_J=\Phi_0 I_c / 2 \pi$ and the charging energy $E_c=(2e)^2/C$, with $I_c$ and $C$ the critical current and capacitance for each of the two nominally identical junctions in each qubit (the first and the third junctions in Fig.~\ref{fig1}b, from top to bottom). For three of the measured devices, characterized by a relatively low Josephson to charging energy ratio $E_J/E_c$, we observe a doublet. In two other devices, with larger $E_J/E_c$, we observe no doublet within the precision determined by the intrinsic qubit linewidth. We note that for the three devices where we observe a doublet, the splitting changes over long periods of time; the maximum value of the splitting is indicated in the table. Nevertheless, the splitting was stable over typically a few days, during which reliable data can be extracted using spectroscopy and coherent control.

For the three devices where we observe a doublet, the maximum observed splitting is a significant fraction of the calculated maximum charge modulation $\delta\Delta_{c}$ (see Table~\ref{tab1}). While two-state fluctuations in the offset charge were observed in experiments on single electron tunneling devices, the amplitude is usually small~\cite{Zimmerli_1992_ChargeNoise}. In our experiment, a small fluctuation would lead to a transition energy change significantly smaller than what we observe in the experiment. This suggests that the observed splitting is most likely due to quasiparticle tunneling across the qubit junctions~\cite{Sun_2010,Riste_2013_qpcharge}. Indeed, the transition frequency $\nu_{ge}(\Phi,n_{\t{g1}},n_{\t{g2}},n_{\t{g3}})$ is periodic with periodicity 1 for each $n_{\t{g}i}$, $i=\overline{1,3}$. When quasiparticles are present on the three islands, the gate charges are given by $n_{\t{g}i}=n_{\t{g}i0}+1/2*n_{\t{q}i}$, $i=\overline{1,3}$, with $n_{\t{g}i0}$ a random offset charge corresponding to a slowly varying background of trapped charges and $n_{\t{q}i}$ the number of quasiparticles trapped on island $i$. For most values of the offset charges, a change in the quasiparticle numbers on one or more islands will induce a change in $\Delta$ comparable with the maximum modulation $\delta\Delta_{c}$.

\begin{figure}[!]
\includegraphics[width=3.4in]{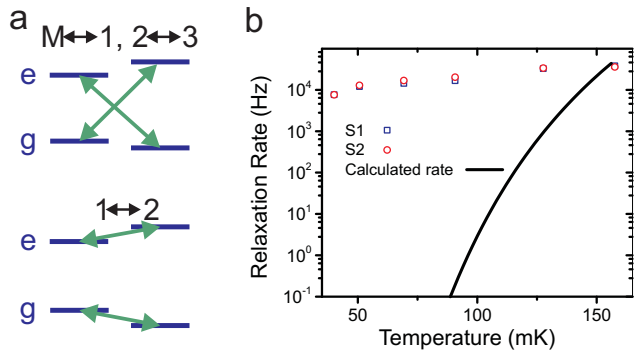}
\caption{\label{fig4}  \textbf{(a)} Representation of the allowed transitions for quasiparticles through different junctions, in the low energy approximation. The other transitions are suppressed due to destructive intereference between quasiparticle and quasihole amplitudes. \textbf{(b)} TSF transition rate vs temperature extracted by excitation at lines S1 (squares) and S2 (circles), and a calculation of the rate for tunneling resulting in a change of the number of quasiparticles on islands 1 and 2.}
\end{figure}

Next we discuss a model for the dynamics of quasiparticle tunneling in the PCQ. Due to the large size of island 3, the changes in energy induced by changes in $n_{\t{g}3}$, and hence by the change of quasiparticle number on this island, are negligible. Therefore, we only consider the dynamics of $n_{\t{q}1}$ and $n_{\t{q}2}$ in the following, connected with quasiparticle tunneling events through junctions M1, 12, and 23 (see Fig.~\ref{fig1}b). We performed numerical calculations of the rates of quasiparticle tunneling events $i\alpha \rightarrow j\beta$, with $i$ and $j$ the initial/final conductor occupied by the quasiparticle and $\alpha$/$\beta$ the initial/final state of the qubit ($i,j \in \{M,1,2,3\}$ and $\alpha,\beta \in \{g,e\}$). We find that in the approximation of low energy quasiparticles the allowed transitions at the qubit symmetry point ($\Phi=\Phi_0/2$) are as indicated in Fig.~\ref{fig4}a: the only allowed processes are those accompanied by a change in qubit energy for M1 and 13 tunneling and those that maintain the qubit energy for 12 tunneling. The same type of selection rules were predicted in~\cite{Leppakangas_2012_FragilityFluxQubits} for a PCQ with three Josephson junctions.

Based on the selection rules for tunneling processes, we consider the block formed by islands 1 and 2 separately from the rest of the circuit. Within this block, fast exchanges of quasiparticles can take place. We find that a quasiparticle in this block undergoes transitions between the islands at a rate in excess of 100 MHz, assuming that the quasiparticle energy does not exceed the superconducting gap by more than 10\%. This assumption on the energy is reasonable given the temperature at which experiments are performed. Quasiparticles transitions within this block are fast and therefore one peak is observed in spectroscopy due to motional narrowing~\cite{Li_2013_MotionalAverage}. Tunneling of quasiparticles between this block and the neighbouring conductors (M and 3) is a slow process which is observed spectroscopically. The measured TSF rate is therefore attributable to tunneling of quasiparticles between the block formed by island 1 and 2 and the rest of the circuit. The thermal equilibrium tunneling rate of quasiparticles between the block 12 and the neighboring junctions is plotted in Fig.~\ref{fig4}b, assuming a superconducting gap of 220~$\mu$eV. While at zero temperature transitions between block 12 and the neighbouring islands are forbidden, the rate is finite at finite temperatures due to the energy dependence of coherence factors~\cite{lutchyn_2005_QpDecayRate,catelani_2012_DecohQpTunnel,Ansari_2012_EnvEffectQpCoupling}, resulting in a finite value of the transition rate (see Supplementary information material).

While at low temperatures the calculated rate is significantly lower than the measured rate, at high temperatures the calculated rate is in reasonable agreement with the measured rate. This indicates the presence of nonequilibrium quasiparticles in the circuit, as previously identified in phase, charge, and transmon qubits ~\cite{martinis_2009_qps,shaw_2008_Qp,Sun_2010,Riste_2013_qpcharge} and recently in a fluxonium qubit~\cite{Pop_2014_CohSupprQuasiparticles}. At the lowest temperatures, the measured value of the qubit energy relaxation time at the symmetry point,  $T_1=4.6\, \mu \t s$, allows extracting an upper bound for the density of non-equilibrium quasiparticles $n_{\t{qp}}<0.7 \mu \t m^{-3}$. The measured quasiparticle tunneling rate is consistent with this density of quasiparticles and an effective temperature of quasiparticles in the range $120-140\,\t{mK}$ (see Supplementary information material). Optimization of decoherence of superconducting flux qubits will have to be addressed both through large values of the $E_J/E_c$ ratio, to reduce dephasing, and through suitable shielding techniques to reduce the nonequilibrium quasiparticle density~\cite{Barends_2011_QpMinimize,Barends_2011_QpMinimize,Saira_2012_VanishingQuasiparticles}, to reduce the role of quasiparticles in energy relaxation.

We have presented experiments in which we extract the time scales associated with a two-state fluctuator coupled to a flux qubit. Using the correlation function of readout results, we extract the transition rates of the TSF over a wide temperature range. We conclude that the source of these fluctuations is tunneling of non-equilibrium quasiparticles, a source of decoherence previously unexplored experimentally for persistent current qubits. These results demonstrate the importance of considering the role of quasiparticles in decoherence of superconducting qubits and will stimulate future theoretical and experimental work on understanding the dynamics of non-equilibrium quasiparticles in complex, multiple-island, superconducting devices.

We thank Chunqing Deng and Florian Ong for help with experiments and useful comments and Frank Wilhelm for valuable discussions. We acknowledge
support from NSERC through Discovery and RTI grants, Canada Foundation for Innovation, Ontario Ministry of Research and Innovation, and
Industry Canada. AL was supported during this work by a Sloan Research Fellowship.



%

\clearpage
\pagebreak
\widetext
\begin{center}
\textbf{\large Supplemental Materials: Dynamics of parametric fluctuations induced by quasiparticle tunneling in superconducting flux qubits}
\end{center}
\setcounter{equation}{0}
\setcounter{figure}{0}
\setcounter{table}{0}
\setcounter{page}{1}
\makeatletter
\renewcommand{\theequation}{S\arabic{equation}}
\renewcommand{\thefigure}{S\arabic{figure}}
\renewcommand{\bibnumfmt}[1]{[S#1]}
\renewcommand{\citenumfont}[1]{S#1}

\section{Introduction}

In this supplementary information material we present calculations of quasiparticle tunneling rates in a persistent current qubit. These results are used in the analysis of the experimental results presented in the main text.

\section{Formulation of the tunneling problem}

We consider a Josephson junction, which is part of the qubit, with electrodes denoted by L(left) and R(right). The tunneling of electrons through this junction is described by the usual transfer Hamiltonian~\cite{bardeen_1961_1}:
\be \label{eq:Htransfer}
H_T=\sum_{p p' \sigma}\l( t_{pp'} c\dg_{L,p \sigma} c_{R,p' \sigma} + t^*_{pp'} c\dg_{R,p' \sigma} c_{L,p \sigma} \r),
\ee
where $p$ and $p'$ are indices for the single particle state quasimomentum, $\sigma=\pm 1$ is a spin index,  $ c_{L,p \sigma} \l( c_{R,p' \sigma '} \r)$ is the annihilation operator for an electron with orbital index $p$($p'$) and spin index $\sigma$ ($\sigma'$) in electrode L(R), and $t_{p p'}$ a transition matrix element for single particle tunneling.

\begin{figure}[h]
\includegraphics[width=2.0in]{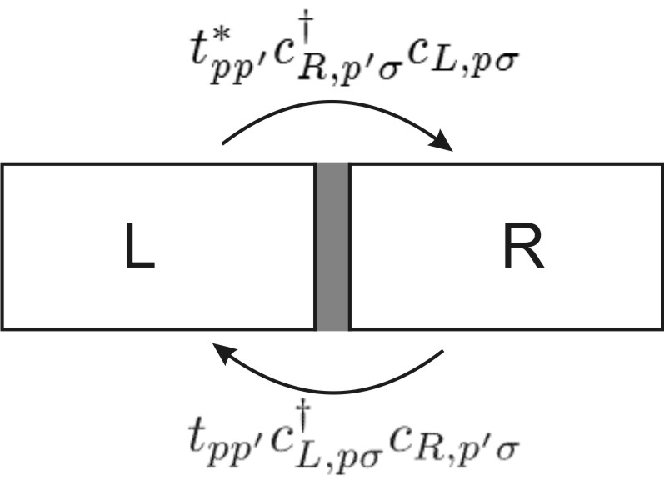}
\caption{\label{figS1} Schematic representation of a Josephson junction. The two arrows indicate different single charge transfer processes, with the operators corresponding to Eq.~\ref{eq:Htransfer} in the text.}
\end{figure}

Since the electrodes are in the superconducting state, it is convenient to express the single particle creation and annihilation operators in terms of quasiparticle operators. We have~\cite{Schrieffer_1999_TheorySuperconductivity,lutchyn_2007_PhDKineticSuperCircuits}:
\be
c_{\alpha,p \sigma}=u_{\alpha, p}\gamma_{\alpha, p \sigma}+\sigma v_{\alpha, p} R_{\alpha} \gamma_{\alpha, -p -\sigma}\dg,
\ee
with $\gamma_{\alpha, p \sigma}$ a quasiparticle annihilation operator and $R_{\alpha}$ a pair annihilation operator for electrode $\alpha$ ($\alpha=L\, \t{or} \,R$). The factors $u_{\alpha, p}$ and $v_{\alpha, p}$ are given by
\be \label{eq:uBogoliubov}
u_{\alpha, p} = \rt{ \frac{1}{2} \l( 1 + \t{sign} \l( \xi_{\alpha, p} \r) \frac{\rt{E_{\alpha,p}^2-\Delta_{\alpha}^2}}{E_{\alpha,p}}\r ) }
\ee
and
\be \label{eq:vBogoliubov}
v_{\alpha, p} = \rt{ \frac{1}{2} \l( 1 - \t{sign} \l( \xi_{\alpha, p} \r) \frac{\rt{E_{\alpha,p}^2-\Delta_{\alpha}^2}}{E_{\alpha,p}}\r ) },
\ee
with $\xi_{\alpha, p} $ the single-electron energy referred to the chemical potential and $\Delta_\alpha$ the superconducting gap for electrode $\alpha$.

We consider a tunneling process in which one quasiparticle is removed from lead $R$ and one quasiparticle is created in lead $L$, corresponding to the operator $\gamma_{L, p \sigma}\dg \gamma_{R, p \sigma}$. This type of process results from the coherent addition of two terms: 1)the removal of one electron from R and the addition of one electron to L, and 2)the removal of one electron from R and the addition of one electron to L combined with the transfer of one Cooper pair. These terms transfer \textbf{charge} in opposite directions, and result from the two types of terms, $c\dg_{L,p \sigma} c_{R,p' \sigma}$ and $c\dg_{R,p' \sigma} c_{L,p \sigma} $ respectively, present in the Hamiltonian \ref{eq:Htransfer}. We are interested in the rate of this process, assuming that the initial state of the qubit (before quasiparticle transfer) is $\ket i$ whereas the final state (after quasiparticle transfer) is $\ket f$. The rate for this process is given by
\begin{widetext}
\be
\Gamma^{i\rightarrow f}_{R \rightarrow L} = \frac{4\pi}{\hbar} \int d E_L \int d E_R D_{\t{qp},L} (E_L) D_{\t{qp},R} (E_R)  f_R(E_R) \l(1-f_L(E_L) \r) \delta ( E_L+\hbar \overline{\omega}_{if} - E_R)\abs{M_{i\rightarrow f}^{R\rightarrow L}}^2 .
\ee
\end{widetext}
In this expression $D_{\t{qp},\alpha} (E_\alpha)$ is the density of states and $f_\alpha(E_\alpha)$ is the probability of occupation for quasiparticles at energy $E_\alpha$ in electrode $\alpha$. The qubit energy change $\hbar \overline{\omega}_{if}=\hbar \omega_{f}-\hbar \omega_{i}$ with $\hbar \omega_{i (f)}$ the initial (final) qubit energy. Finally, the matrix element
\be
\abs{M_{i\rightarrow f}^{R\rightarrow L}}^2=\overline{\abs{t}^2}  \abs{m_{i\rightarrow f}^{R\rightarrow L}}^2
\ee
with $\overline{\abs{t}^2}$ the average value of the coupling element $t_{pp'}$ in the coupling Hamiltonian~\ref{eq:Htransfer} and
\be \label{eq:mif}
m_{i\rightarrow f}^{R\rightarrow L}  =  u_{L, p}^*  u_{R, p'}  \bra{f} O_1^{R \rightarrow L} \ket{i} - v_{L, -p}  v_{R, -p'}^*  \bra{f} O_2^{R \rightarrow L} \ket{i}.
\ee

In Eq.~\ref{eq:mif}, the operators $O_1$ and $O_2$ act on qubit states in the following way. The initial state of the qubit is represented in the charge basis as
\be \label{eq:ichargerep}
\ket{i}=\sum_{c_L,c_R}a_i(c_L,c_R)\ket{c_L,q_L;c_R,q_R}.
\ee
This state is a superposition of states with different numbers of Cooper pairs, $c_L$ and $c_R$,  on the electrodes $L$ and $R$ respectively. The integers $q_L$ and $q_R$ are fixed integers representing the number of unpaired quasiparticles on electrodes $L$ and $R$ respectively. The number of unpaired quasiparticles is conserved under the action of the mesoscopic Hamiltonian of the qubit, consisting of charging energy and Josephson tunneling terms. Similaryly to \ref{eq:ichargerep}, the final state, following the tunneling of a quasiparticle through the tunnel junction between $L$ and $R$, is given by
\be \label{eq:ichargerep}
\ket{f}=\sum_{c_L,c_R}a_f(c_L,c_R)\ket{c_L,q_L+\delta q;c_R,q_R-\delta q}.
\ee
Here $\delta q$ is the change in the number of unpaired quasiparticles; $\delta q=1$ ($-1$) for a $R\rightarrow L$ ($L\rightarrow R$) quasiparticle tunneling event. The matrix elements of the operators $O_1$ and $O_2$ for $R \rightarrow L$ tunneling are given by:
\be
\bra{f} O_1^{R \rightarrow L} \ket{i} = \sum_{c_L,c_R} a_f(c_L,c_R)^* a_i(c_L,c_R),
\ee
and
\be
\bra{f} O_2^{R \rightarrow L} \ket{i} = \sum_{c_L,c_R} a_f(c_L-1,c_R+1)^* a_i(c_L,c_R).
\ee

Based on \ref{eq:mif}, \ref{eq:uBogoliubov}, and \ref{eq:vBogoliubov} we can write
\be
\abs{m_{i\rightarrow f}^{R\rightarrow L}}^2 = \abs{A_{1,if}}^2 + \abs{A_{2,if}}^2 - 2\frac{\Delta_L \Delta_R}{E_L E_R}\t{Re}\l( A_{1,if} A_{2,if}^* \r),
\ee
where we introduced $A_{k,if}=\bra{f} O_k^{R \rightarrow L} \ket{i}$, with $k \in \{ 1,2 \}$ and $i,f \in \{ g,e\}$ the initial and final states of the qubit respectively.

\section{Selection rules for the flux qubit}
We calculate the matrix elements $A_{k,if}$ numerically. We start by setting up a circuit model for a PCQ with Josephson junctions which generalizes the model of Orlando \emph{et al.}~\cite{Sorlando_1999_1}. We include gate charges coupled to islands 1, 2, and 3 (see Fig.1 in the main text) to represent both random offset charges and unpaired charges due to quasiparticles. The Hamiltonian is represented in the charge basis and the eigenvalues/eigenvectors are determined by numerical diagonalization. The matrix elements $A_{k,if}=\bra{f} O_k^{R \rightarrow L} \ket{i}$ are then calculated numerically. For a PCQ biased at the symmetry point, and considering tunneling between islands M and 1 or 2 and 3 (see Fig. 1b in the main text) we find the following selection rules:
\begin{itemize}
\item $A_{1,gg}=A_{2,gg}$ and $A_{1,ee}=A_{2,ee}$. We denote these values by $m_{\t{c},gg}$ and $m_{\t{c},ee}$ respectively.
\item $A_{1,ge}=-A_{2,ge}$ and $A_{1,eg}=-A_{2,eg}$. We denote these values by $m_{\t{nc},ge}$ and $m_{\t{nc},eg}$ respectively.
\end{itemize}
These transition rules are shown schematically in Fig. 4a of the main text.

With these selection rules, the transition rates can be expressed as follows. For transitions in which the qubit remains in a state with the same energy index, $\alpha=g$ or $e$, the transition rate is given by
\begin{widetext}
\be \label{eq:generalrateconserving}
\Gamma^{\alpha\rightarrow \alpha}_{R \rightarrow L} = \frac{2 G_T}{e^2} \abs{m_{\t{c}, \alpha \alpha}}^2  \int d E_L \int d E_R\;  \delta ( E_L+\hbar \overline{\omega}_{\alpha \alpha} - E_R)    d_{\t{qp},L} (E_L) d_{\t{qp},R} (E_R)  f_R(E_R) \l(1-f_L(E_L) \r)    \l( 1- \frac{\Delta_L \Delta_R}{E_L E_R} \r),
\ee
\end{widetext}
where we introduced the normal state tunnel conductance $G_T$ of the junction and the normalized quasiparticle density of states
\be
d_{\t{qp},\alpha} (E_\alpha)  =  \frac {E_\alpha}   {\rt{E_{\alpha}^2-\Delta_{\alpha}^2}}
\ee
for the two electrodes ($\alpha=L,R$).

For transitions in which the qubit changes the energy brach from $\alpha$ to $\beta$ ($\alpha \neq \beta$), we have
\begin{widetext}
\be \label{eq:generalratenonconserving}
\Gamma^{\alpha\rightarrow \beta}_{R \rightarrow L} = \frac{2 G_T}{e^2} \abs{m_{\t{nc}, \alpha \beta}}^2  \int d E_L \int d E_R \; \delta ( E_L+\hbar \overline{\omega}_{\alpha \beta} - E_R)    d_{\t{qp},L} (E_L) d_{\t{qp},R} (E_R)  f_R(E_R) \l(1-f_L(E_L) \r)  \l( 1+ \frac{\Delta_L \Delta_R}{E_L E_R} \r).
\ee
\end{widetext}

\section{Transitions rates for different types of processes}
In this section we present the expressions for the different quasiparticles tunneling rates, corresponding to the different combinations of initial/final states of the qubit.
\subsection{Processes for $e\rightarrow g$ qubit transitions}
The conservation law for this type of process is given by
\be
E_R=E_L+\hbar \overline{\omega}_{eg}.
\ee
Here $E_R$ is the energy of the quasiparticle in lead $R$, which tunnels into a state with energy $E_L$ in lead $L$. The qubit changes its energy by an amount
\be
\hbar \overline{\omega}_{eg}=E_{qb,g}^{\t{initial parity}}-E_{qb,e}^{\t{final parity}}.
\ee
We used the superscript to indicate that the energy of the qubit depends on the parity (determined by the number of quasiparticles). We have
\be
\hbar \overline{\omega}_{if}\approx E_{qb,f}^{\t{initial parity}}-E_{qb,i}^{\t{initial parity}} \approx E_{qb,f}^{\t{final parity}}-E_{qb,i}^{\t{final parity}},
\ee
which holds because the qubit energy level splitting is much larger than the modulation of each energy by changes in parity, for any value of the offset charges (as discussed in the main text).

\begin{figure*}[!]
\includegraphics[width=6.5in]{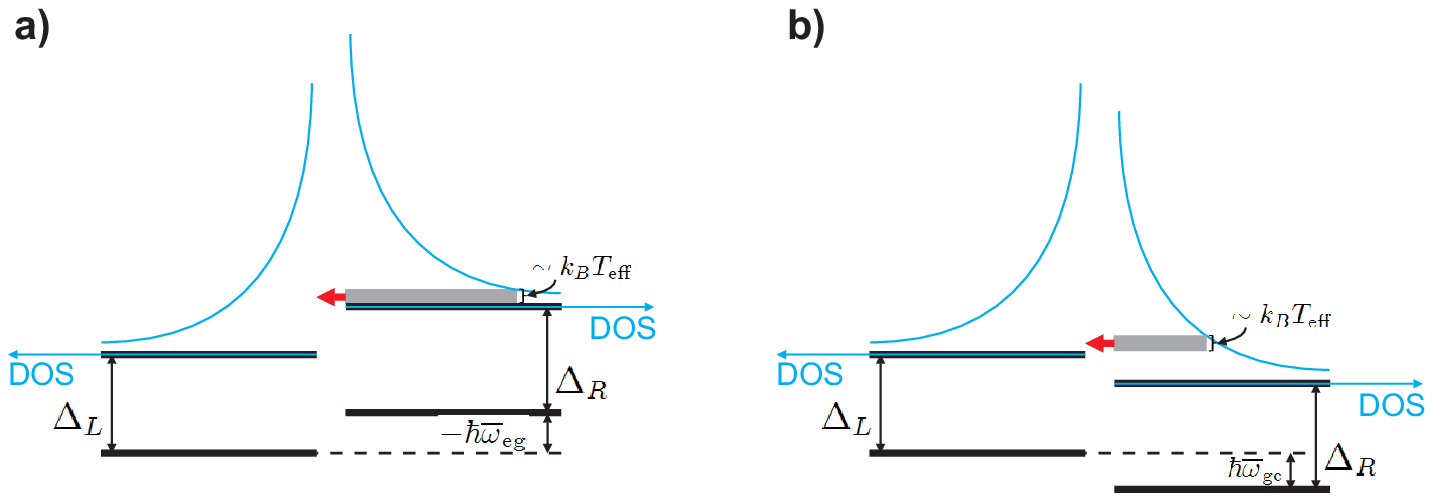}
\caption{\label{figS2} Representation of quasiparticle tunneling for $e\rightarrow g$ (panel a) and $g\rightarrow e$ (panel b) processes.}
\end{figure*}

The expression for the transition rate for the $e\rightarrow g$ process is given by
\begin{widetext}
\be
\Gamma^{e\rightarrow g}_{R \rightarrow L} = \frac{2G_T}{e^2} \abs{m_{\t{nc}, \alpha \beta}}^2    \int_{\max \{\Delta_R, \hbar \overline{\omega}_{eg} + \Delta_L \}}^\infty d E_R \;   \frac {\l(E_R - \hbar \overline{\omega}_{eg} \r)E_R + \Delta_R \Delta_L}   {\rt{   \l[ \l( E_R-\hbar \overline{\omega}_{eg} \r)^2 -\Delta_L^2\r]  \l[ E_R^2 -\Delta_R^2\r]    }}   f_R(E_R) \l(1-f_L(E_R-\hbar \overline{\omega}_{eg}) \r) .
\ee
\end{widetext}
(Note that $\overline{\omega}_{eg}$ is a negative quantity.)
As illustrated in Fig.~\ref{figS2}a, this rate results from the tunneling of quasiparticles which occupy energies just above the superconducting gap in electrode $R$ to states which are most likely empty into electrode $L$. The states in electrode $L$ which are occupied after tunneling takes place are at energies $\Delta_R-\Delta_L+\hbar \overline{\omega}_{ge}$ above the superconducting gap, where, assuming that the qubit energy level splitting is large compared to the imbalance between the superconducting gaps in the two electrodes, the density of states does not have any singularity. Therefore it is possibly to factor out of the integral a contribution which is proportional to the density of quasiparticles in the electrode $R$, $n_{\t{qp},R}$, given by:
\be
n_{\t{qp},R} = 4 \int_{\Delta_R}^\infty  d E_R \frac {E_R} {\rt{E_R^2-\Delta_R^2}} \overline{D}_R (E_{F,R}) f_R(E_R),
\ee
where $\overline{D}_R (E_{F,R})$ is the energy density of states, normalized to volume, in electrode $R$, at the Fermi energy $E_{F,R}$. With the approximation $\l(1-f_L(E_R-\hbar \overline{\omega}_{ge}) \r) \approx 1$, we obtain

\begin{widetext}
\be \label{eq:Gammaegwithnqp}
\Gamma^{e\rightarrow g}_{R \rightarrow L} =  \frac{G_T}{2e^2} \abs{m_{\t{nc}, \alpha \beta}}^2 \frac{n_{qp,R}}  {\overline{D}_R (E_{F,R})} \rt{ \frac{\Delta_R - \hbar \overline{\omega}_{eg}+\Delta_L}  {\Delta_R - \hbar \overline{\omega}_{eg} - \Delta_L}  }.
\ee
\end{widetext}
This relation holds when the superconducting gap values $\Delta_L$ and $\Delta_R$, as well as $\Delta_R-\Delta_L+\hbar \overline{\omega}_{ge}$, are significantly larger than the effective temperature of the quasiparticle distributions.

\subsection{Processes for $g\rightarrow e$ qubit transitions}
\label{subseq:gtoe}

The configuration of levels for this case is illustrated by the diagram in Fig.~\ref{figS2}b. If the temperature is low enough, states in the electrode $L$ are mostly unoccupied. The total rate of tunneling from $R$ to $L$ is reduced, with respect to the $g\rightarrow e$ case, due to the fact that only quasiparticles with energy $\Delta_L-\Delta_R+\hbar \overline{\omega}_{ge}\approx \hbar \overline{\omega}_{ge}$ above the gap can tunnel out of $R$. The transition rate depends, in this case, on the details of the distribution of the quasiparticles over energy, and not only on the total density. Qualitatively, we can argue that the rate of this process is given by an expression of the form
\be \label{eq:gtoedetailedbalance}
\Gamma^{g\rightarrow e}_{R \rightarrow L} = \Gamma^{e\rightarrow g}_{R \rightarrow L}e^{\frac{-\hbar \omega_{ge}}{k_B T_{\t{eff}}}}.
\ee
Here we assume that the quasiparticle distribution, which may be in general a nonequilibrium distribution, is a Fermi distribution with the effective temperature $T_{\t{eff}}$. The relation~\ref{eq:gtoedetailedbalance} can be understood as a detailed balance condition: the ratio of the rates $\Gamma^{g\rightarrow e}_{R \rightarrow L}/\Gamma^{e\rightarrow g}_{R \rightarrow L}$ is equal to the ratio of probabilities of occupation of the excited and ground states of the qubit, which is the Boltzman factor with a temperature which corresponds to the environment.

\subsection{Processes for $g\rightarrow g$ qubit transitions}

We start with expression~\ref{eq:generalrateconserving} and we assume for definiteness $\hbar \overline{\omega}_{gg}>0$ and also we take $\Delta_L=\Delta_R\equiv \Delta$. We use, in addition, $f_R(E_R)\approx e^{-E_R/k_BT}$ and $1-f_L(E_L)\approx 1$ (which are justified as long as the temperature is significantly below the superconducting gap). The expression for the transition rate is
\begin{widetext}
\be
\Gamma^{g\rightarrow g}_{R \rightarrow L} = \frac{2G_T}{e^2} \abs{m_{\t{c}, gg}}^2    \int_{\Delta}^\infty d E_L \; \frac  {E_L \l( E_L+\hbar \overline{\omega}_{gg} \r) - \Delta^2}   {\rt{ \l(E_L^2-\Delta^2\r)  \l( \l(E_L+\hbar \overline{\omega}_{gg} \r)^2-\Delta^2   \r)   }} e^{-\frac{E_L+\hbar \overline{\omega}_{gg}}{k_B T}}.
\ee
\end{widetext}
By using $\Delta \gg k_B T$ and $\hbar \overline{\omega}_{gg} \ll k_B T$, which are both justified for typical experimental conditions, we obtain
\be \label{eq:Gammaggfinal}
\Gamma^{g\rightarrow g}_{R \rightarrow L} = \frac{2G_T}{e^2} \abs{m_{\t{c}, gg}}^2    k_B T e^{-\frac{\Delta}{k_B T}}.
\ee
We assumed $\hbar \overline{\omega}_{gg}>0$ to start with. If we assume instead $\hbar \overline{\omega}_{gg}<0$, the same result is obtained as long as $\abs{\hbar \overline{\omega}_{gg}}\ll k_B T$.

\subsection{Processes for $e\rightarrow e$ qubit transitions}

The calculations proceed in a fashion fully similar with those for the $g\rightarrow g$ in the previous section. With similar assumptions, namely $\abs{\hbar \overline{\omega}_{ee}}\ll k_B T$, $\Delta \gg k_B T$, and $\Delta_L=\Delta_R\equiv \Delta$, we find
\be \label{Gammaeefinal}
\Gamma^{e\rightarrow e}_{R \rightarrow L} = \frac{2G_T}{e^2} \abs{m_{\t{c}, ee}}^2    k_B T e^{-\frac{\Delta}{k_B T}}.
\ee

\subsection{Accuracy of the approximate expressions for the transition rates}

We checked expressions \ref{eq:Gammaegwithnqp}, \ref{eq:gtoedetailedbalance},\ref{eq:Gammaggfinal}, and \ref{Gammaeefinal} for the case of a thermal distribution of quasiparticles, for the entire temperature range explored in the experiments (40 to 158 mK) against the rates calculated by numerical integration in Mathematica using relations \ref{eq:generalrateconserving} and \ref{eq:generalratenonconserving}. The agreement is within 5\%.

\section{Application of the theory to the experiments}
The energy relaxation time of the qubit, measured at its symmetry point ($\Phi=\Phi_0/2$), is $T_1=4.6\,\mu\t{s}$ for the device for which detailed measurements were presented in the main text. Other measurements on similar devices produced energy relaxation times ranging from a few microseconds to 10 microseconds at the symmetry point. In all of these cases, the calculated energy relaxation induced by the resonator to which the qubit is coupled, through the Purcell effect~\cite{Houck_2008_SpontanousEmissionTransmon}, was a negligible contribution to the measured rate. We therefore believe that quasiparticle tunneling is a substantial contribution to the energy relaxation rate. Based on \ref{eq:Gammaegwithnqp} we can place an upper bound on the quasiparticle density. We sum up the rates corresponding to quasiparticle tunneling through both the $M1$ and $3M$ junctions, which give the most important contributions to energy relaxation. Using the numerically determined value for $m_{\t{nc},eg}$, and assuming a superconducting gap energy $\Delta=220\,\mu\t{eV}$, we find that the quasiparticle density $n_{\t{qp}}<0.7\,\mu\t{m}^{-3}$.

\begin{figure}[!]
\includegraphics[width=3.4in]{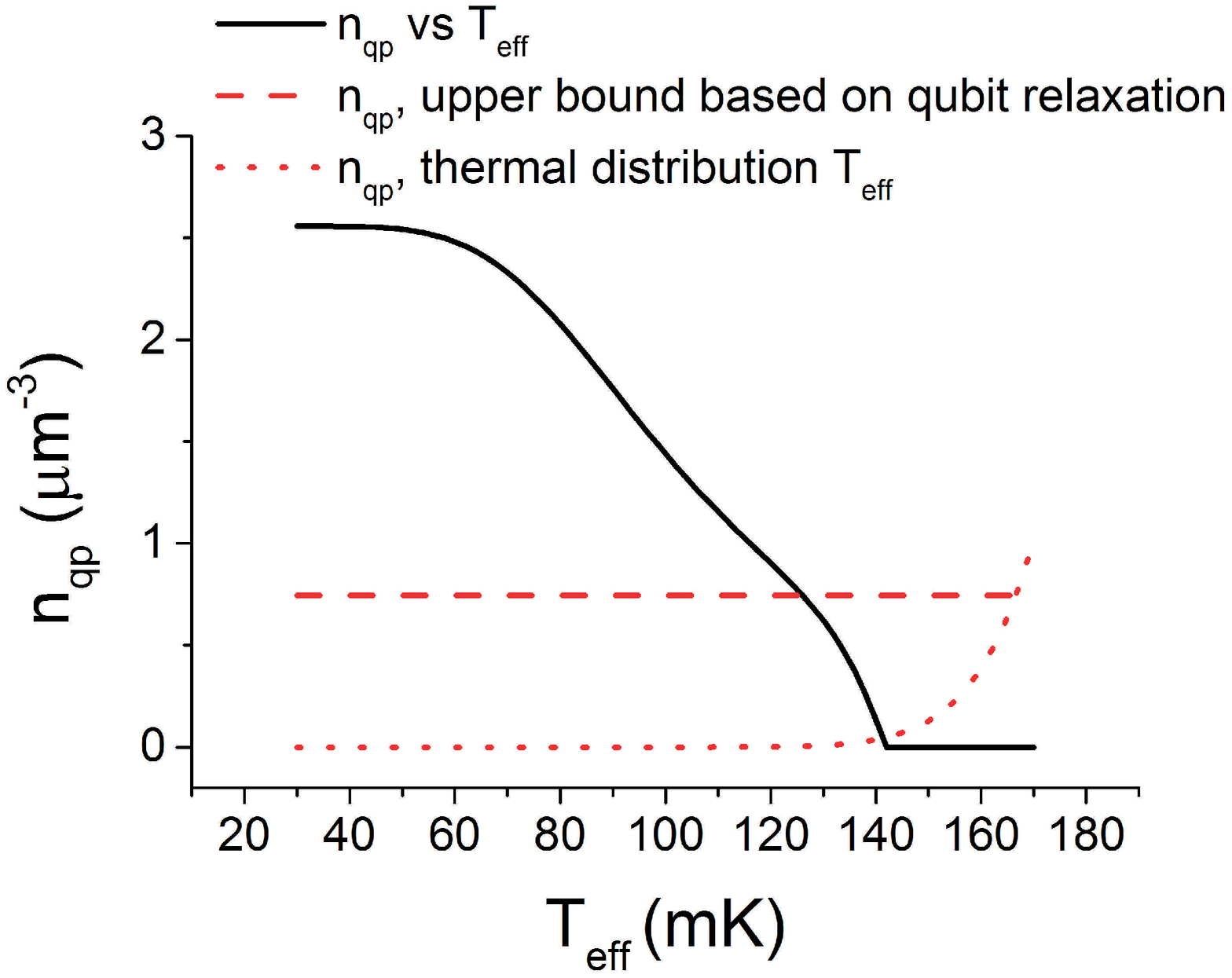}
\caption{\label{figS3} Relation between the quasiparticle density $n_{qp}$ and the effective temperature $T_{\t{eff}}$, corresponding to the experimentally measured parity change rate at 40 mK (black continuous line), the upper bound on quasiparticle density based on the qubit energy relaxation time (red dashed line) and the density for a thermal distribution at temperature $T_{\t{eff}}$ (red dotted line). }
\end{figure}

Next, we verify the consistency between the transition rate between the two parity configurations, denoted $S1$ and $S2$ in the main paper, measured at the lowest temperature of 40 mK, and the rate model developed in this paper. The measured rate in the correlation measurement is the sum of the transitions between the two states of different parity, given by
\be
\Gamma_c = P_g (\Gamma^{g\rightarrow g} + \Gamma^{g\rightarrow e}) + P_e (\Gamma^{e\rightarrow g} + \Gamma^{e\rightarrow e}),
\ee
where each rate $\Gamma^{\alpha \rightarrow \beta}$ (with $\alpha,\beta=g\,\t{or}\,e$) is a sum over all the processes in which the qubit undergoes a transition from state $\alpha$ to state $\beta$ ($\alpha,\beta = g\,\t{or}\,e$), accompanied by an observable change of parity. The probabilities $P_g$ ($P_e$) are the probabilities for the qubit to be in the $g$($e$) state. They are determined based on the histogram of readout results, with the qubit prepared by a waiting time significantly longer than the energy relaxation time (as shown in Fig. 2d in the main text). At the qubit symmetry point we find $P_e=1\%$. The rate corresponding to qubit energy relaxation ($e\rightarrow g$) depends only on the quasiparticle density (see Eq.~\ref{eq:Gammaegwithnqp}), with the assumptions discussed above. The other rates depend on the effective temperature of the quasiparticles (see equations \ref{eq:gtoedetailedbalance},\ref{eq:Gammaggfinal}, and \ref{Gammaeefinal}), as discussed in sub-section~\ref{subseq:gtoe}. With $\Gamma_c$ a function of both the quasiparticle density $n_{\t{qp}}$ and the effective temperature $T_{\t{eff}}$, the measured value of $\Gamma_c$ only provides a relation between $n_{\t{qp}}$ and $T_{\t{eff}}$. In Fig.~\ref{figS3} we show the quasiparticle density versus the effective temperature, together with the upper bound on the density determined above based on the qubit relaxation time. For reference, we also show the thermal distribution corresponding to the temperature $T_{\t{eff}}$. This result shows that the observed value of the transition rate can be explained if we assume an effective temperature ranging approximately between 120 and 140 mK.


\end{document}